# Magnetocaloric effect and critical behavior near the paramagnetic to ferrimagnetic phase transition temperature in $TbCo_{2-x}Fe_x$

Madhumita Halder, S. M. Yusuf,* and M. D. Mukadam

*Solid State Physics Division, Bhabha Atomic Research Centre, Mumbai 400085, India*

K. Shashikala

*Chemistry Division, Bhabha Atomic Research Centre, Mumbai 400085, India*

Magnetocaloric effect (MCE) in $TbCo_{2-x}Fe_x$ has been studied by dc magnetization measurements. On substituting Fe in $TbCo_2$, not only the magnetic transition temperature is tuned to room temperature, but also the operating temperature range for MCE is increased from 50 K for $TbCo_2$ to 95 K for $TbCo_{1.9}Fe_{0.1}$. The maximum magnetic entropy change ($-\Delta S_M$) for $TbCo_{1.9}Fe_{0.1}$ is found to be 3.7 J $kg^{-1}$ $K^{-1}$ for a 5 T field change, making it a promising candidate for magnetic refrigeration near room temperature. The temperature dependent neutron diffraction study shows a structural phase transition (from cubic to rhombohedral phase with lowering of temperature) which is associated with the magnetic phase transition and these transitions broaden on Fe substitution. To investigate the nature of the paramagnetic to ferrimagnetic phase transition, we performed a critical exponent study. From the derived values of critical exponents, we conclude that $TbCo_2$ belongs to the 3D Heisenberg class with short-range interaction, while on Fe substitution it tends towards mean-field with long-range interaction. The derived values of critical exponents represent the phenomenological universal curve for the field dependence of $\Delta S_M$, indicating that $TbCo_2$ and $TbCo_{1.9}Fe_{0.1}$ belong to two different universality classes.

PACS number(s): 75.30.Sg, 61.05.fm, 75.40.Cx

* Electronic mail: smyusuf@barc.gov.in FAX: +91 22 25505151

# I. INTRODUCTION

Cubic Laves phase compounds $R\text{Co}_2$, where $R$ is a rare earth, have become potential candidates for magnetocaloric effect (MCE) as they show a large change in magnetic entropy ($\Delta S_M$ ) around their magnetic ordering temperatures.[1] They are of great interest since their discovery because of their interesting magnetic and electronic properties.[2-5] These compounds with the light rare earth elements like Pr, Nd, Sm, Gd and Tb are reported to show second order magnetic phase transitions while, with heavy rare earth elements like Dy, Ho and Er, first order metamagnetic transitions are reported.[6] It is known that in these $R\text{Co}_2$ compounds, the magnetic moments of the Co atoms align antiparallel to those of the rare earth atoms.[2] In such compounds, the observed magnetic moments of the $R$ atoms are close to their theoretically expected large moment values from their free ionic states.[7] Hence, a large MCE is expected for $R\text{Co}_2$ compounds. The magnetic state of Co depends on the induced moment at Co site due to 4*f*-3*d* exchange interactions and also on the lattice parameter in the $R\text{Co}_2$ compounds.[6] Here, we have investigated the effect of substitution of Fe at the Co site in $\text{TbCo}_2$ on the MCE. Some other alloys and compounds which show large MCE, are $\text{Gd}_5\text{Si}_2\text{Ge}_2$ alloy,[8] $\text{MnFe}(\text{P}_{1-x}\text{As}_x)$,[9] $\text{La}(\text{Fe}_{13-x}\text{Si}_x)$,[10] Ni-Mn-Ga Heusler alloys,[11-13] perovskites,[14-16] *etc.* However, most of these materials undergo a first order magnetic phase transition. Even though the change in magnetic entropy is large in such type of materials, they exhibit large thermal and field hystereses on variation of magnetization with temperature and magnetic field, respectively. So, their usage for MCE applications is limited. Moreover, for a material showing first order magnetic phase transition, the value of $-\Delta S_M$ is highest near the magnetic transition temperature and falls rapidly with temperature, making its usage limited over a narrow temperature range. From the practical application point of view, materials with a large MCE over a broad temperature range are desired.[17-20]

For $TbCo_2$ compound, the paramagnetic to ferrimagnetic second order phase transition temperature $T_C$ (231 K) lies well below the room temperature.[1,6] It is also known to undergo a structural transition near $T_C$. Below $T_C$, it has a rhombohedral crystal structure with a space group $R\overline{3}m$ and above $T_C$ it has a cubic crystal structure with a space group $Fd\overline{3}m$.[21,22] In the present study, we have shown that an appropriate (Fe) chemical substitution not only tunes the paramagnetic to ferrimagnetic transition temperature to around room temperature but also increases the operating temperature range of magnetic refrigeration without significant reduction in the $\Delta S_M$ value. We have also performed the temperature dependent neutron diffraction study on $TbCo_{2-x}Fe_x$ to investigate the effect of Fe substitution on the structural phase transition and correlated that with the MCE. Generally, a structural phase transition is associated with a first order magnetic phase transition,[23-25] but $TbCo_2$ is reported to undergo a second order magnetic phase transition.[26,27] This further motivated us to investigate the nature of the order of magnetic phase transition in detail in these $TbCo_{2-x}Fe_x$ compounds by carrying out dc magnetization and neutron diffraction studies. Fe substitution at the Co site in $TbCo_2$ also provides an opportunity to investigate whether this transition can be tuned by a quenched disorder i.e. by chemical substitution. A critical exponent analysis in the vicinity of the magnetic phase transition is a powerful tool to investigate the details of the magnetic interaction responsible for the transition.[28] There are reports in literature that a quenched disorder can tune a first order phase transition to a second order phase transition.[29,30] However, no experimental study of the critical phenomena has been reported yet on $TbCo_2$ or $TbCo_{2-x}Fe_x$. The present critical exponent study deals with a case where the effect of quenched disorder is investigated for a second order magnetic phase transition. The nature and dimensionality of the magnetic interaction responsible for the transition have also been brought out from the present critical exponent study. It is also shown that $\Delta S_M$ ($H$) curves are unique for each universality class.

## II. EXPERIMENTAL DETAILS

The polycrystalline $TbCo_{2-x}Fe_x$ samples ($x$ = 0, 0.06, 0.08 and 0.1) with constituent elements of 99.9 % purity, were prepared by an arc-melting under argon atmosphere. An excess of Tb (3 wt %) was added to the starting compositions to prevent the formation of Co-rich phases. For better chemical homogeneity, the samples were re-melted many times. After melting, they were annealed in vacuum-sealed quartz tube at 850 ºC for 7 days. The powder x-ray diffraction (XRD) using the Cu-$K_\alpha$ radiation in the 2θ range of 10 – 90º with a step of 0.02º was carried out on all the samples at room temperature. Magnetization ($M$) measurements were carried out using a commercial (Oxford Instruments) vibrating sample magnetometer. For dc magnetization measurements, the samples were field-cooled (FC) under 0.02 T field down to 5 K. The measurements were then carried out in the warming cycle from 5 to 320K under the same field. The magnetization isotherms were recorded at various temperatures with an interval of 5 K up to a maximum applied magnetic field of 5 T. For critical exponent study, magnetization isotherms were measured for $x$ = 0 and 0.1 samples up to a maximum field of 5 T over a temperature range of $T_C \pm 10$ K with an interval of 1 K. Neutron diffraction patterns were recorded at various temperatures for $x$ = 0 and 0.1 samples using the powder diffractometer – II (λ = 1.249 Å) at the Dhruva Research Reactor, Trombay, Mumbai, India. Differential scanning calorimetry (DSC) measurements were carried out using Mettler Toledo DSC 822, with an empty aluminum pan as a reference. The measurements were carried out in the heating as well as cooling cycles with a scanning rate 5 K/min for the $x$ = 0 and 0.1 samples.

## III. RESULTS AND DISCUSSION

XRD patterns at room temperature confirmed that the samples had no trace of any impurity phase. The FC magnetization versus temperature curves are shown in Fig. 1 for all samples. The paramagnetic to ferrimagnetic transition temperature $T_C$ is derived from the minima of the d$M$/d$T$ vs $T$ curves. $T_C$ has been found to increase with Fe substitution (231, 275, 290 and 303 K for $x = 0$, 0.06, 0.08 and 0.1, respectively). The observed increase in $T_C$ on substitution of Fe at the Co site is due to an enhanced exchange interaction between the 3$d$ transition metal ions.[31] The increase in $T_C$ is associated with the density of states at the Fermi level and on the splitting energies of the 3$d$ sub bands.[32,33]

### A. Magnetocaloric effect

The change in entropy ($S$) of a magnetic material on applying magnetic field ($H$) is related to the change in magnetization ($M$) with respect to temperature ($T$) and it can be expressed by the following Maxwell relation.

$$\left(\frac{\partial S}{\partial H}\right)_T = \left(\frac{\partial M}{\partial T}\right)_H \tag{1}$$

The magnetic entropy change $\Delta S_M$ can be calculated from the magnetization isotherms as follows

$$\Delta S_M(H,T) = \int_0^H \left(\frac{\partial M(H,T)}{\partial T}\right)_H dH \tag{2}$$

For magnetization measurements made at discrete temperature interval and fields, $\Delta S_M$ ($H, T$) can be approximated as

$$\Delta S_M(H,T) = \sum_i \frac{M_{i+1}(T_{i+1},H) - M_i(T_i,H)}{T_{i+1} - T_i}\Delta H \tag{3}$$

Fig. 2 shows a series of magnetization isotherms measured at various temperatures for the samples with $x$ = 0, 0.06 and 0.1. Here the hysteresis effect is negligibly small. Fig. 3 shows the variation of -$\Delta S_M$ with temperature for the $x$ = 0, 0.06 and 0.1 samples. The maxima in the -$\Delta S_M$ vs $T$ curves are found to be around $T_C$ and -$\Delta S_M$ increases with the increase in the applied magnetic field. The peak of the -$\Delta S_M$ ($T$) curve also shifts towards room temperature on Fe substitution. Another useful parameter which decides the efficiency of a magnetocaloric material is the relative cooling power (RCP) or the refrigerant capacity. It is the heat transfer between the hot and the cold reservoirs during an ideal refrigeration cycle. This represents numerically the area under the -$\Delta S_M$ vs $T$ curve. It can be calculated by integrating the -$\Delta S_M$ ($T$) curve over the full width at half maximum,[34]

$$RCP = \int_{T_{Cold}}^{T_{Hot}} \Delta S_M \, dT \qquad (4)$$

Here, $T_{Hot}$ and $T_{Cold}$ are higher and lower temperatures at half maximum of the -$\Delta S_M$ ($T$) peak and can be considered as the temperatures of the hot and cold reservoirs, respectively. The inset of Fig. 3 marks the area under the -$\Delta S_M$ ($T$) curve corresponding to RCP for the $x$ = 0.06 sample. The -$\Delta S_M$, $T_{Cold}$, $T_{Hot}$, and RCP values of various compositions at field values of 1 and 5 T are given in Table 1. For $x$ = 0.1 sample, due to experimental limitation we could not perform magnetization measurements above 320 K. So we extrapolated the -$\Delta S_M$ ($T$) curve to derive RCP. It can be seen (from Fig. 3) that by substituting Fe at the Co site, the width of the -$\Delta S_M$ ($T$) curve increases. This is because the magnetic transition spreads over a wide temperature range. This may be attributed to the presence of some magnetic randomness (disorder) at low temperatures in the Tb sublattice on substituting Fe at the Co site. A similar kind of broadening was reported in $TbNi_2$ on substituting Fe at the Ni site by Singh *et al* [35] which was due to the magnetic randomness in the Tb sublattice. It was shown that this randomness occurred from the Tb sublattice and not from the Ni or Fe sublattices. For the $x$ = 0.06 sample, the operating temperature range (defined as a difference between $T_{Hot}$ and $T_{Cold}$) is over 216 - 305 K i.e. ~ 89 K at $\Delta\mu_0H$ = 5 T and for the $x$ = 0.1 sample, the operating temperature range is over 245 - 340 K i.e. ~ 95 K, while for the $x$ = 0 sample it was only 50 K temperature range. The values of $-\Delta S_M$ and the operating temperature range for the present $x$ = 0.1 sample and other reported materials showing giant MCE close to room temperature are given in Table 2. The operating temperature range for the $x$ = 0.1 sample is much broader than that for other reported giant MCE materials in the same temperature range (see Table 2). Such a broad operating temperature range is important for practical applications.

Thus, by substituting Fe in $TbCo_2$, we not only tuned the $T_C$ from 231 K towards room temperature but also increased the operating temperature range for the MCE. To further increase the operating temperature range, composites of these alloys with different Fe concentrations can be made. A series of such compositions can be used in a cascaded way for magnetic cooling over a broad temperature range. Another important factor to be considered for magnetic refrigerant material is the hysteresis loss. It has been already pointed out that the hysteresis in the $M(\mu_0H)$ curves (shown in Fig. 2) in the operating temperature range is negligible for the Fe substituted samples, thus increasing the working efficiency of the material. This is important for a reversibility of the MCE of refrigerant materials.

### B. Neutron Diffraction Study

We performed the neutron diffraction study on the parent sample as well as on the $x$ = 0.1 sample to find whether the structural phase transition is coupled with the magnetic phase transition. Neutron diffraction patterns were measured at various temperatures in magnetically ordered and paramagnetic states for both $x$ = 0 and 0.1 samples. The patterns measured at 22 K and 300 K for both $x = 0$ and 0.1 samples are shown in Fig. 4. The diffraction patterns were analyzed by the Rietveld refinement technique using the FULLPROF program.[36] The starting values of the atomic positions and lattice constants for the parent sample were taken from literature.[22] The analysis of the room temperature diffraction pattern for the $x = 0$ sample reveals that it has cubic crystal structure with a space group $Fd\bar{3}m$. The refinement of the diffraction pattern at 22 K shows that the low temperature phase is rhombohedral with a space group $R\bar{3}m$. This is in agreement with the neutron[22,37] and X- ray powder diffraction[21] results reported earlier. The refinement of the diffraction pattern at 22 K for $x = 0.1$ sample reveals that the low temperature phase is rhombohedral, whereas at 300 K there is a co-existence of both rhombohedral and cubic phases. From the analysis of the diffraction patterns measured at various temperatures we find that near the magnetic transition temperature $T_C$, there is also a structural phase transition. For both the samples, we find that there is an increase in the *a*-axis lattice parameter and, a decrease in the *c*-axis lattice parameter with increasing temperature [Fig. 5 (a)]. A relatively larger change in the lattice parameters is found near the magnetic transition temperature, indicating that the structural phase transition is coupled with the magnetic transition. Fig. 5 (b) shows the variation of the "unit cell" volume with temperature for $x = 0$ and 0.1 samples. For the cubic unit cell, the lattice constant $a$ is used to calculate the unit cell volume ($a^3$). For the rhombohedral phase, the quantities $\sqrt{2}a$ and $c/\sqrt{3}$ ($a$ and $c$ are the lattice parameters of the rhombohedral phase) are equivalent to the cubic lattice. These quantities are, therefore, used to calculate the "unit cell" volume of the rhombohedral phase. The continuous change in the "unit cell" volume across the phase transition temperatures indicates that the structural phase transition is of second order in nature. There is an increase in the average Co-Co bond length on substituting Fe at the Co site, shown in the inset of Fig. 5 (b). For the $x = 0.1$ sample, the structural transition also shifts towards a higher temperature along with the magnetic transition temperature. The structural phase transition occurs near the same temperature as that of the magnetic transition temperature, indicating that there is a strong magneto-elastic coupling for both $x = 0$ and 0.1 samples. Here, a quantitative phase analysis of the rhombohedral and cubic phases has been made as a function of temperature across the structural/magnetic (magnetostructural) phase transition for both $x = 0$ and 0.1 samples and the derived phase fractions are plotted in Fig. 6. For the $x = 0.1$ sample, with increasing temperature, we find that the growth of the cubic phase (at the expense of the rhombohedral phase) spreads over a wide temperature range (260 K to beyond room temperature) indicating that the structural transition is broadened on Fe substitution. This is in good agreement with the results obtained from the dc magnetization studies, where we observe that the width of the $-\Delta S_M$ $(T)$ curve increases on Fe substitution. The second order nature of the structural phase transition can be due to the soft phonons associated with the structural phase transition.[38,39] Thus the structural transition coupled with the magnetic transition enhances the magnetocaloric effect. The refinement of the neutron diffraction pattern at 22 K for the $x = 0$ sample shows a ferrimagnetic ordering of the Tb and Co moments along the *c*-axis of the rhombohedral structure as reported in literature.[22] Tb has an ordered moment of 8.49 ± 0.17 $\mu_B$ per Tb ion, while Co has an ordered moment of 1.07 ± 0.02 $\mu_B$ per Co ion which are consistent with the values reported earlier.[22] The saturation moment is calculated to be 6.34 $\mu_B$/f.u. For the $x = 0.1$ sample, the Rietveld refinement of the neutron diffraction pattern at 22 K shows that there is a ferrimagnetic ordering of Tb with Co and Fe. Tb has an ordered moment of 8.44 ± 0.02 $\mu_B$ per Tb ion, Co has an ordered moment of 1.12 ± 0.01 $\mu_B$ per Co ion, while Fe has an ordered moment of 1.84 ± 0.03 $\mu_B$ per Fe ion. The saturation moment is calculated to be 6.13 $\mu_B$/f.u. Figs. 7 (a) and (b) show variation of magnetic moment as a function of temperature for both $x = 0$ and 0.1 samples.

## C. Critical Behavior Study

To analyze the nature of the magnetic phase transition in detail, we have carried out critical exponent study near the Curie temperature $T_C$ for the $x = 0$ and 0.1 samples. According to the scaling hypothesis,[40] for a second order phase transition around $T_C$, the critical exponents β (associated with the spontaneous magnetization), γ (associated with the initial susceptibility), and δ (associated with the magnetization isotherm) are related as

$$M_S(T) = M_0(-\varepsilon)^{\beta},\ \varepsilon < 0 \qquad (5)$$

$$\chi_0^{-1}(T) = (h_0 / M_0)\varepsilon^{\gamma},\ \varepsilon > 0 \qquad (6)$$

At $T_C$, the exponent δ relates magnetization $M$ and applied magnetic field $H$ by

$$M = DH^{1/\delta},\ \varepsilon = 0 \qquad (7)$$

Here $M_S$ ($T$) is the spontaneous magnetization; $\chi_0^{-1}(T)$ is the inverse of susceptibility; ε is the reduced temperature $(T\text{-}T_C)/T_C$; $M_0$, $h_0/M_0$, and $D$ are the critical amplitudes. According to the scaling hypothesis,[41] $M(H,\varepsilon)$ is a universal function of $T$ and $H$ and the experimental $M(H)$ curves are expected to collapse into the universal curve

$$M(H,\varepsilon) = \varepsilon^{\beta} f_{\pm}(H / \varepsilon^{\beta+\gamma}) \qquad (8)$$

with two branches, one for temperatures above $T_C$ and the other for temperatures below $T_C$. Here $f_+$ for $T>T_C$ and $f_-$ for $T<T_C$, are regular functions.

Fig. 8 (a) shows the Arrott plot ($M^2$ vs $H/M$) for the $x = 0$ sample. Only a positive slope of the $M^2$ vs $H/M$ has been observed, indicating that the paramagnetic to ferrimagnetic transition is of second order according to the Banerjee criteria.[42] According to the mean field theory, such curves should give a series of straight lines for different temperatures and the isotherm at $T = T_C$ should pass through the origin.[40] But in our case, we find that the curves of the Arrott plot are nonlinear showing that the mean field theory is not applicable in this case. Therefore, we tried to analyze the data using the modified Arrott plots (MAPs), based on the Arrott-Noakes equation of state.[43] According to the MAPs method, the $M^{1/\beta}$ versus $(H/M)^{1/\gamma}$ plots at various temperatures are parallel to each other at high magnetic fields. The critical isotherm at $T = T_C$ is the line which passes through the origin. The initial values of β and γ are chosen in such a way that they give straight lines in the MAPs. The values of spontaneous magnetization $M_S(T)$ and $\chi_0^{-1}(T)$ are obtained from a linear extrapolation of MAPs at fields above 0.1 T to the intercept with the $M^{1/\beta}$ and $(H/M)^{1/\gamma}$ axes, respectively. Only the high field linear region is used for the analysis, because MAPs deviate from linearity at low field due to the mutually misaligned magnetic domains.[44] The values of $M_S(T)$ and $\chi_0^{-1}(T)$ are plotted as a function of temperature. By fitting these plots with Eqs. 5 and 6, we get the new values of β and γ. By using these new values of β and γ, new MAPs are constructed. In an iterative process, we get the stable values of β, γ, and $T_C$. In Fig. 9 (a) the $M_S(T)$ and $\chi_0^{-1}(T)$ verses temperature curves are plotted. The continuous curve shows the power law fits of Eqs. (5) and (6) to $M_S(T)$ and $\chi_0^{-1}(T)$, respectively. Eq. (5) gives the value of β = 0.380(4) with $T_C$ = 224.83(7) K and Eq. (6) gives the value of γ = 1.407(8) with $T_C$ = 224.67(1) K for the $x$ = 0 sample. These values of β and γ are close to the values predicted for a short-range Heisenberg model (β = 0.368 and γ = 1.396).[45] To find the value of δ, the $M_S$ (225 K, $H$) versus $H$ isotherm is plotted on the log-log scale [inset of Fig. 9 (a)]. According to Eq. (7), this should be a straight line with a slope 1/ δ. From the linear fit of the straight line, we obtain δ is derived to be 4.85(3). The value of δ can also be calculated from the Widom scaling relation[46]

$$\delta = 1 + \gamma / \beta \qquad (9)$$

The calculated value of δ is 4.70. The value of δ, calculated from the magnetization isotherm, is in agreement with the scaling hypothesis within the experimental errors.The critical exponent analysis can be justified by the $M\varepsilon^{-\beta}$ vs $H\varepsilon^{-(\beta+\gamma)}$ plot. According to Eq. 8 all the data should fall on one of the two curves. The scaled data are plotted on a log scale as shown in Fig.11 (a). It can be clearly seen that all the data fall on either of the two branches of the universal curve, one for temperatures above $T_C$ and the other for temperatures below $T_C$.

Now for the $x$ = 0.1 sample, Fig. 8 (b) shows the Arrott plot or $M^2$ vs $H/M$ curves. From the Arrott plot we can find that the curves are almost linear, suggesting that it is close to the mean field theory. By choosing initial values of β and γ, we did the same analysis for $x$ = 0.1 sample. In Fig.9 (b), the $M_S$ ($T$) and $\chi_0^{-1}(T)$ verses temperature curves for $x$ = 0.1 sample are shown. Eq. (5) gives the value of β = 0.541(1) with $T_C$ = 304.48(7) K and Eq. (6) gives the value of γ = 1.023(2) with $T_C$ = 304.27(3) K for the $x$ = 0.1 sample. These values of β and γ are close to the values predicted from mean-field theory (β = 0.5 and γ = 1).[40] The value of δ obtained from the fitting of the $M_S$ (304 K, $H$) versus $H$ isotherm plotted on the log-log scale, as shown in inset of Fig. 9 (b) is 2.75(4). From the Widom scaling relation the value of δ is calculated to be 2.89, which is in agreement with value of δ derived from the magnetization isotherm within the experimental errors. $M\varepsilon^{-\beta}$ vs $H\varepsilon^{-(\beta+\gamma)}$ plot on a log scale is shown in Fig.10 (b). It can be clearly seen that all the data fall on either of the two branches of the universal curve, one for temperatures above $T_C$ and the other for temperatures below $T_C$. Physically, β describes how the ordered moment grows below $T_C$. Smaller the value of β, faster is the growth of the ordered moment. γ describes the divergence of the magnetic susceptibility at $T_C$. Smaller values yield a sharper divergence, and δ describes the curvature of $M$ ($H$) at $T_C$, with smaller values reflecting less curvature and slower saturation. The value of β increases for the $x$ = 0.1 sample reflecting a slower growth of the ordered moment with decreasing temperature. This is consistent with the neutron diffraction study (Fig. 6), where we found that the magnetostructural transition broadens on Fe substitution, which results in broadening of the width of the $\Delta S_M$ ($T$) curve (Fig.3), indicating a slow magnetic phase transition on Fe substitution. On substituting Fe, the value of δ decreases, indicating a less curvature and a slower saturation in the $M$ ($H$) curves. This can be observed from the $M$ vs $H$ curves of both $x$ = 0 and 0.1 samples (shown in Fig. 2).

A phenomenological universal curve for the field dependence of $\Delta S_M$ has been proposed by Franco *et al*.[47] The major assumption is based on the fact that if an universal curve exists, then the equivalent points of the $\Delta S_M$ ($T$) curves measured at different applied fields should collapse onto the same universal curve. It is also shown that these curves are unique for each universality class.[48] The MCE data of different materials of same universality class should fall onto the same curve, irrespective of the applied magnetic field. The field dependence of $\Delta S_M$ is given by the following equation

$$\Delta S_M \Big|_{T=T_C} \propto H^n \quad \text{where } n = 1+1/\delta(1-1/\beta) \qquad (10)$$

Using the values of β and δ obtained above, the values of $n$ are calculated to be 0.653 and 0.707 for $x$ = 0 and 0.1samples, respectively. The values of $n$ obtained from fitting of Eq. (10) are 0.65(1) and 0.76(2) for $x$ = 0 and 0.1 samples, respectively, which are in good agreement with that obtained from the critical exponents using MAPs. From Fig. 11, it is clear that $x$ = 0 and 0.1 samples belong to two different universality classes. This shows that for materials with same universality class, there exists an universal curve for the magnetic entropy change. This existence of a universal curve helps us to predict the response of a particular material under different experimental conditions and can be useful for designing materials for practical application.

In general a structural transition is associated with a first order magnetic transition.[23-25] In case of $TbCo_2$, we find that there is a structural phase transition coupled with a magnetic phase transition but the nature of the magnetic phase transition is of second order. Magnetostructural transition with second order has been observed earlier also.[49] The second order nature of the structural phase transition can be due to the soft phonon modes.[38,39] To further confirm the nature of the magnetostructural phase transition, we have performed the differential scanning calorimetry (DSC) measurements on the $x = 0$ and 0.1 samples in the heating as well as cooling cycles (Figs. 12 (a) and (b)). For the $x = 0$ sample, the DSC scan [Fig. 12 (a) ] shows a phase transition at 230 K close to its magnetic phase transition temperature (231 K) obtained from the magnetization measurements. A negligible thermal hysteresis of $\Delta T < 1$ K (at a scanning rate of 5K/min) is observed between the heating and cooling scans. The transition has a long tail on the lower temperature side indicating a second order nature of the phase transition. For the $x = 0.1$ sample, the DSC scan [Fig. 12(b)] shows a phase transition at 304 K, close to its magnetic phase transition temperature (303 K). Here also a second order phase transition is evident from the observed negligible thermal hysteresis ($\Delta T < 1$ K) and a long tail on the lower temperature side. On substituting Fe, we find that not only the magnetic transition is coupled with the structural transition and the magnetic phase transition remains second order, but the transition spreads over a wide temperature range. On substituting Fe, we find that the value of β increases and value of γ decreases. Change in the values of critical exponents with quenched disorder is a general trend observed in amorphous ferromagnets.[50] There, the change in the values of critical exponents is due to the formation of finite magnetic clusters in the infinite ferromagnetic network. For the $x = 0.1$ sample, the value of β is greater than the value predicted by the mean-field theory. In $Nd_{0.7}Sr_{0.3}MnO_3$ single crystal, the value of β and γ are found to be 0.57(1) and 1.16(3), respectively, which is due to the formation of ferromagnetic clusters above $T_C$.[51] For the $x = 0.1$ sample, the paramagnetic to ferrimagnetic transition takes place over a wide temperature range indicating a possible existence of ferromagnetic clusters.[52] For the $x = 0$ sample, the critical exponents are close to the Heisenberg model, while for $x = 0.1$ sample, the exponents are close to the mean field theory, indicating that there is a crossover from short-range interaction to long-range interaction on substituting Fe. The increase in the value of β and decrease in the value of γ with increasing Fe concentration were also evident in $(Fe_xMn_{1-x})_{75}P_{16}B_6Al_3$ alloys.[53] There, the value of δ decreases and $T_C$ increases with Fe concentration as observed for $TbCo_{2-x}Fe_x$ in the present study. A possible reason for the dominance of long-range interactions could be Ruderman-Kittel-Kasuya-Yosida (RKKY) interactions in these intermetallic compounds. In amorphous ferromagnets, the deviation from the Heisenberg model has been attributed to the RKKY interactions found in such systems.[50] In rare earth based intermetallics, the interaction is mainly of RKKY type. On substituting Fe, the RKKY interactions dominate and the exchange interaction extends beyond the nearest-neighbors leading to a long-range interaction which enhances the exchange interaction and an enhancement in $T_C$ is observed.

## IV. CONCLUSION

Here, we have investigated the effect of Fe substitution at the Co site on the magnetocaloric effect in $TbCo_2$. We observe that the substitution of Fe at the Co site causes an increase in $T_C$, which could be due to an enhancement in the exchange interaction between the 3*d* transition metal ions. For $\Delta\mu_0H$ = 5 T, the maximum reversible magnetic entropy change of 3.7 J kg$^{-1}$ K$^{-1}$ for the $x$ = 0.1 sample, has been observed near the magnetic transition temperature (303 K). Substitution of Fe at the Co site not only tunes the $T_C$ towards room temperature but also causes an increase in the MCE operating temperature range (~95 K for $x$ = 0.1 sample). The magnetostructural transition spreads over a wide temperature. This is confirmed by the temperature dependent neutron diffraction experiments. Thus, the broad operating temperature range with high values of -$\Delta S_M$, and RCP, and negligible hysteresis make $TbCo_{1.9}Fe_{0.1}$ a potential candidate for magnetic refrigeration near room temperature. The temperature dependent neutron diffraction study on the magnetostructural transitions has given an understanding of their intertwined magnetic and structural properties. The results are useful to explore their implications on large magnetocaloric effect shown by these compounds across the magnetostructural phase transition. The DSC measurements show the absence of thermal hysteresis for both the $x = 0$ and 0.1 samples indicating the second order nature of the magnetostructural phase transition. The critical exponents β, γ and δ are obtained using MAPs and Widom scaling relation. The field and temperature dependent magnetization behavior follows the scaling theory and all the data points fall on the two distinct branches, one for $T < T_C$ and the other for $T > T_C$. The values of critical exponents obtained are used to show that the equivalent points of the $\Delta S_M (T)$ curves measured at different applied fields collapse onto the same universal curve i.e. these curves are unique for each universality class. The critical exponents obtained from MAPs agree fairly well with the $\Delta S_M (H)$ curve. For the $x = 0$ sample, the values of critical exponents are close to that of 3D Heisenberg model with short-range interaction. For the $x = 0.1$ sample, the values of critical exponents are close to that of mean-field theory with long-range interaction, indicating that on substituting Fe, there is a crossover from short-range interaction to long-range interaction and an enhancement in $T_C$.

## ACKNOWLEDGEMENT

M. H. acknowledges the help provided by A. B. Shinde and L. Panicker for performing the neutron diffraction experiments and DSC measurements, respectively.

Table 1

The magnetic entropy change -$\Delta S_M$, $T_{Cold}$, $T_{Hot}$, and relative cooling power RCP values of various compositions at field values of 1 and 5 T.

| $TbCo_{2-x}Fe_x$ | $\Delta\mu_0 H$ (T) | -$\Delta S_M$ (J kg$^{-1}$K$^{-1}$) | $T_{Cold}$ (K) | $T_{Hot}$ (K) | RCP (J kg$^{-1}$) |
|---|---|---|---|---|---|
| $x = 0$ | 1 | 2.5 | 217 | 239 | 74 |
| | 5 | 6.9 | 205 | 255 | 357 |
| $x = 0.06$ | 1 | 1.1 | 253 | 286 | 48 |
| | 5 | 3.9 | 216 | 305 | 299 |
| $x = 0.1$ | 1 | 1.0 | 269 | 315 | 46 |
| | 5 | 3.7 | 245 | 340 | 271 |

Table 2

The magnetic entropy change -$\Delta S_M$ and the operating temperature range for the $x = 0.1$ sample and other GMCE materials close to room temperature.

| Materials | $T_C$ (K) | $\Delta\mu_0 H$ (T) | -$\Delta S_M$ (J kg$^{-1}$ K$^{-1}$) | $T_{hot}-T_{cold}$ (K) | Reference |
|---|---|---|---|---|---|
| $TbCo_{1.9}Fe_{0.1}$ | 303 | 5 | 3.7 | 95 | This work |
| Gd | 292 | 5 | 9.5 | 70 | [8] |
| $Gd_5Ge_2Si_2$ | 276 | 5 | 18.6 | 25 | [8] |
| $MnFeP_{0.45}As_{0.55}$ | 300 | 5 | 18.0 | 23 | [9] |
| $Ni_{55}Mn_{20}Ga_{25}$ | 313 | 5 | 86.0 | 1 | [13] |

**List of Figures**

FIG. 1: (Color online) Temperature dependence of normalized magnetization for various compositions at 0.02 T applied field, where $M_{5K}$ is the observed magnetization at 5 K.

FIG. 2: Magnetization isotherms at various temperatures for the $x$ = 0, 0.06 and 0.1 samples. There is negligible hysteresis over the operating temperature range. For clarity, the $M(H)$ curves for both increasing (O) as well as decreasing (+) applied magnetic fields are shown only for $x$ = 0.06 and $x$ = 0.1 samples at 200 and 250 K, respectively.

FIG. 3: (Color online) Magnetic entropy change $-\Delta S_M$ vs temperature for $\Delta\mu_0 H$ = 5 T for $x$ = 0, 0.06 and 0.1 samples. The inset shows the $-\Delta S_M$ vs $T$ curve for $\Delta\mu_0 H$ = 5 Tesla for $x$ = 0.06 sample. The shaded area corresponds to the relative cooling power.

FIG. 4: (Color Online) Neutron diffraction patterns of $x$ = 0 and 0.1 samples at 22 and 300 K. The solid lines represent the Rietveld refined patterns. The difference between the observed and calculated patterns is also shown at the bottom of each curve by solid lines. The vertical bars indicate the position of allowed Bragg peaks.

FIG. 5: (Color Online) Temperature dependence of (a) lattice constants and (b) unit cell volume of the rhombohedral (equivalent to cubic lattice) and cubic phases for the $x$ = 0 and 0.1 samples. The error bars are within the symbol. Inset shows variation of the Co-Co average bond length with temperature for the samples $x$ = 0 and 0.1.

FIG. 6: (Color online) Quantitative phase analysis of the rhombohedral and cubic phases as a function of temperature across the magnetostructural phase transition for both the (a) $x$ = 0 and (b) 0.1 samples. (b) The dashed line above 300 K is the extrapolation of the data obtained below 300 K. Region between the vertical dotted lines shows the temperature range in which both the phases co-exist.

FIG. 7: (Color online) Temperature dependence of magnetic moments (per ion) of (a) Tb and Co in $TbCo_2$ and (b) Tb, Co and Fe in $TbCo_{1.9}Fe_{0.1}$ along the crystallographic *c*-axis. The temperature variation of the net ordered moment values per formula unit of $TbCo_2$ and $TbCo_{1.9}Fe_{0.1}$ are also plotted.

FIG. 8: Isotherms $M^2$ vs $H/M$ at different temperatures close to the Curie temperature for (a) $x = 0$ and (b) $x = 0.1$ samples.

FIG. 9: Temperature dependence of the spontaneous magnetization $M_S$ ($T$) and the inverse initial susceptibility $\chi_0^{-1}(T)$. The solid lines are the fitted curves obtained from Eqs. (5) and (6), respectively for (a) $x = 0$ and (b) $x = 0.1$ samples . The $M$ vs $H$ on a log-log scale at $T$ = 225 K for $x = 0$ sample and at $T$ = 304 K for $x = 0.1$ sample are shown as an inset of (a) and (b), respectively. The straight line is the linear fit with Eq. (7).

FIG. 10: (Color online) Logarithmic scaling plot of $M|\varepsilon|^{-\beta}$ verses $H|\varepsilon|^{-(\beta+\gamma)}$ in the critical region. All the experimental data fall on either of the two branches of the universal curve for (a) $x = 0$ and (b) $x = 0.1$ samples.

FIG. 11: Field dependence of the magnetic entropy change $-\Delta S_M$ for the $x = 0$ and 0.1 samples. The solid lines are the fitted curves using Eq. 10.

FIG. 12: (Color online) Heat flow vs temperature for (a) $x = 0$ and (b) $x = 0.1$ samples in the heating and cooling cycles.

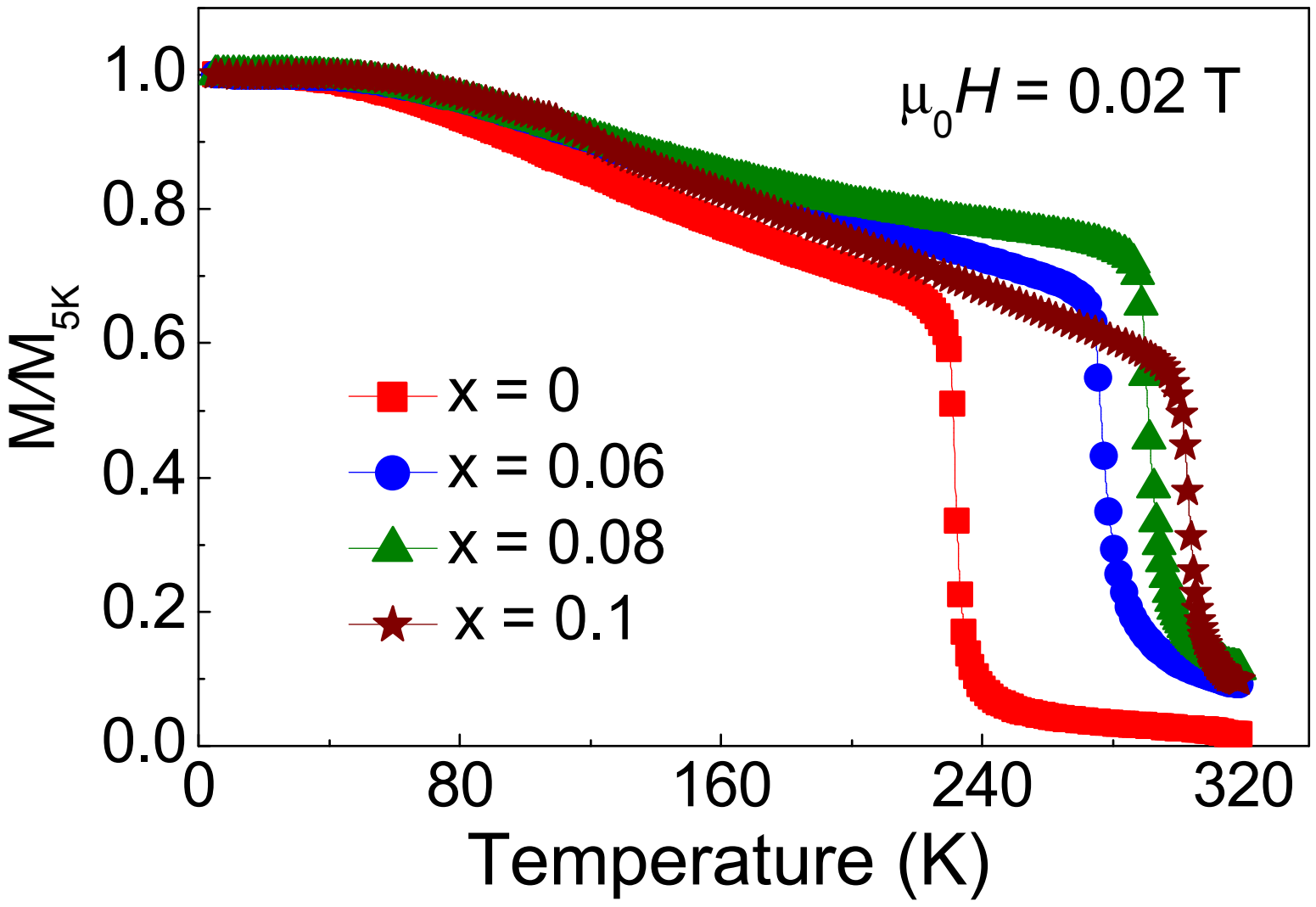

Figure 1
Halder *et al*.

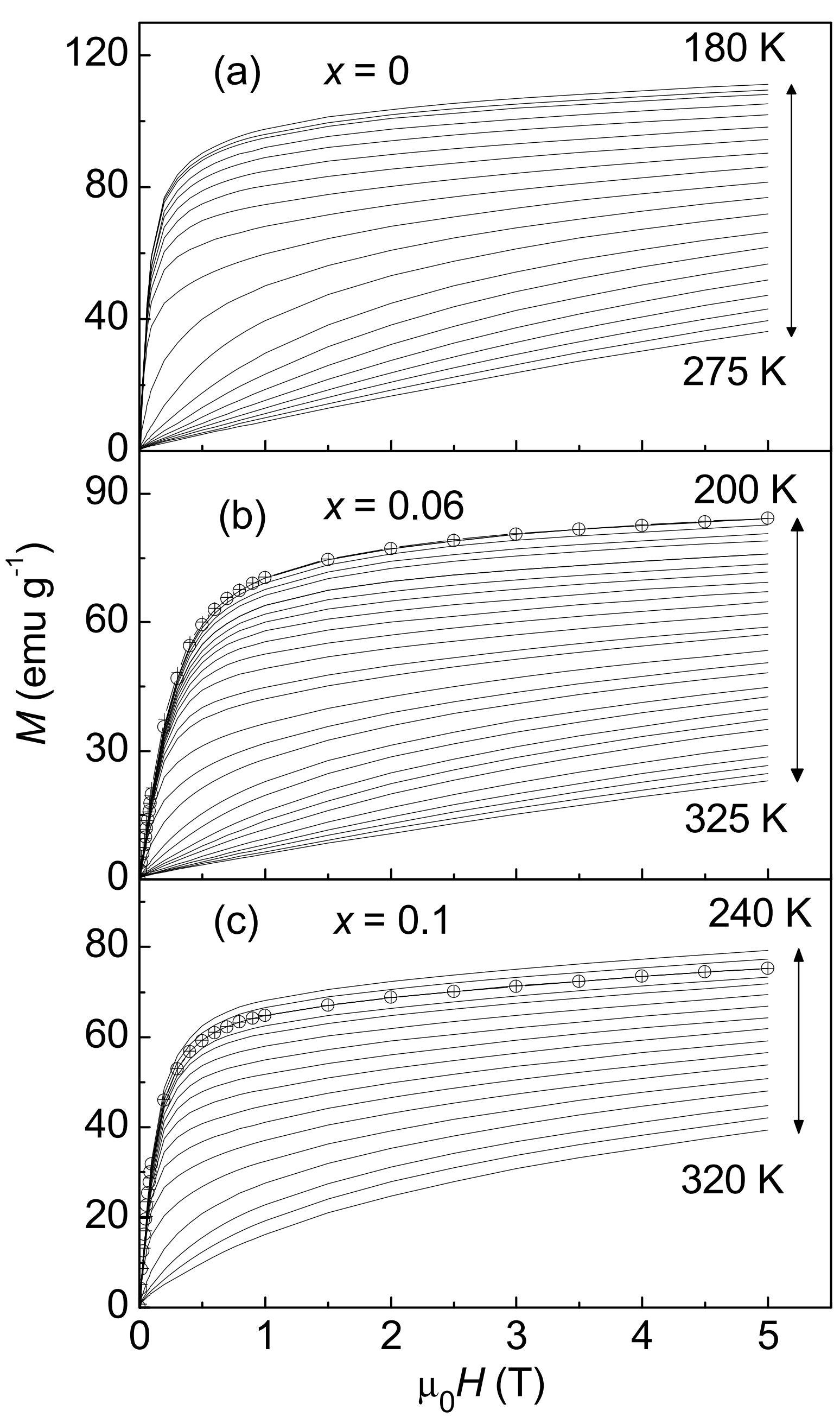

Figure 2
Halder *et al*.

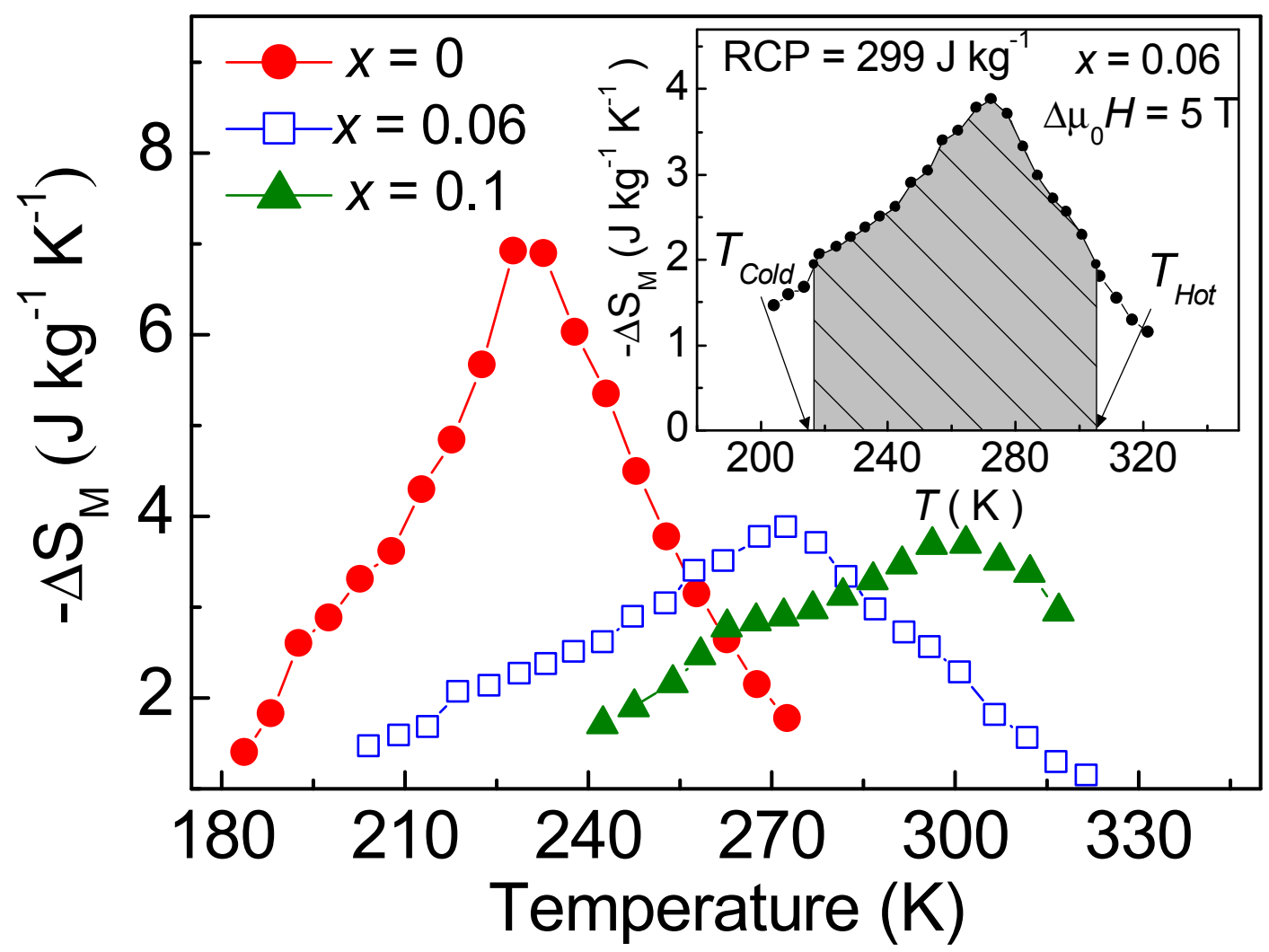

Figure 3
Halder *et al*.

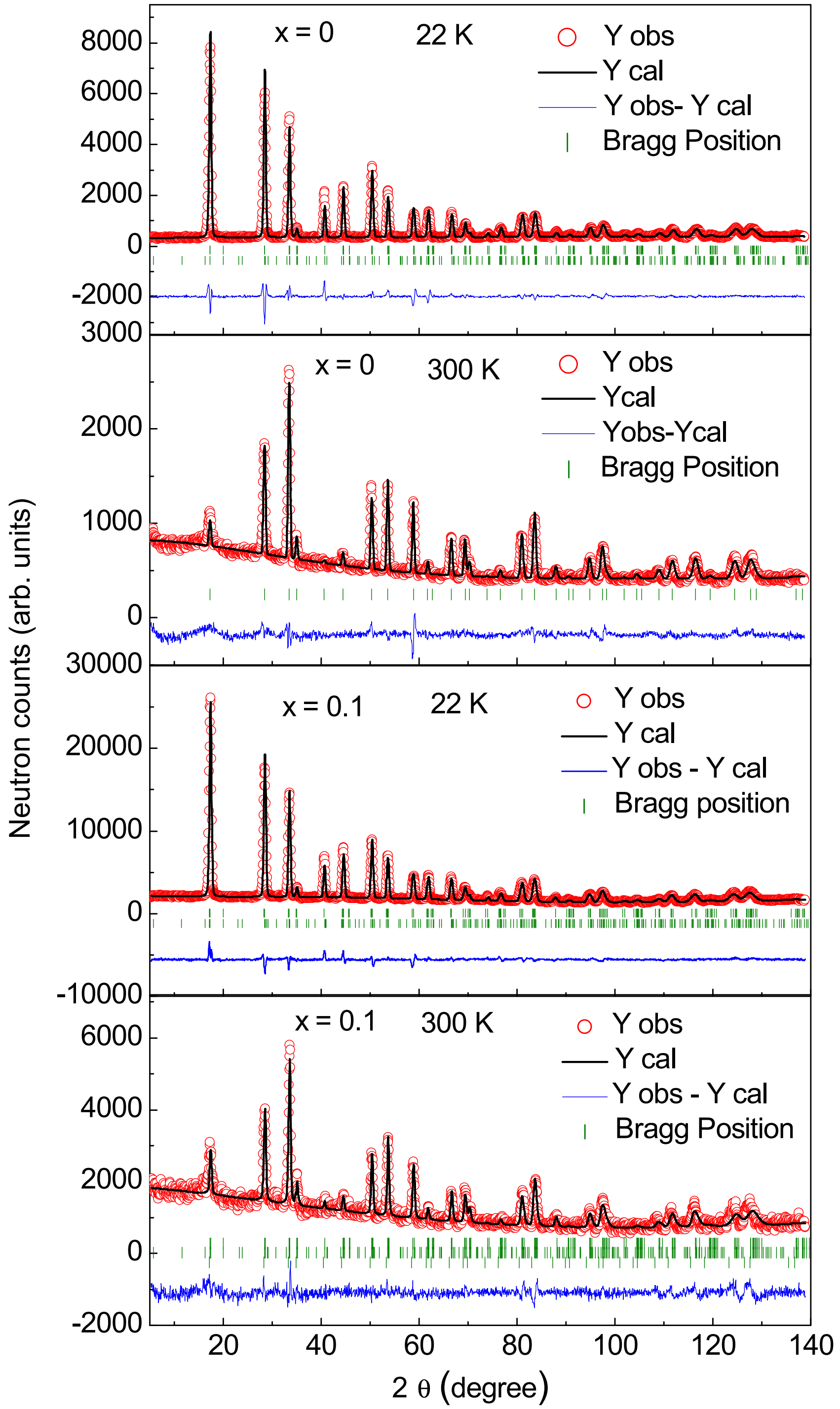

Figure 4
Halder *et al*.

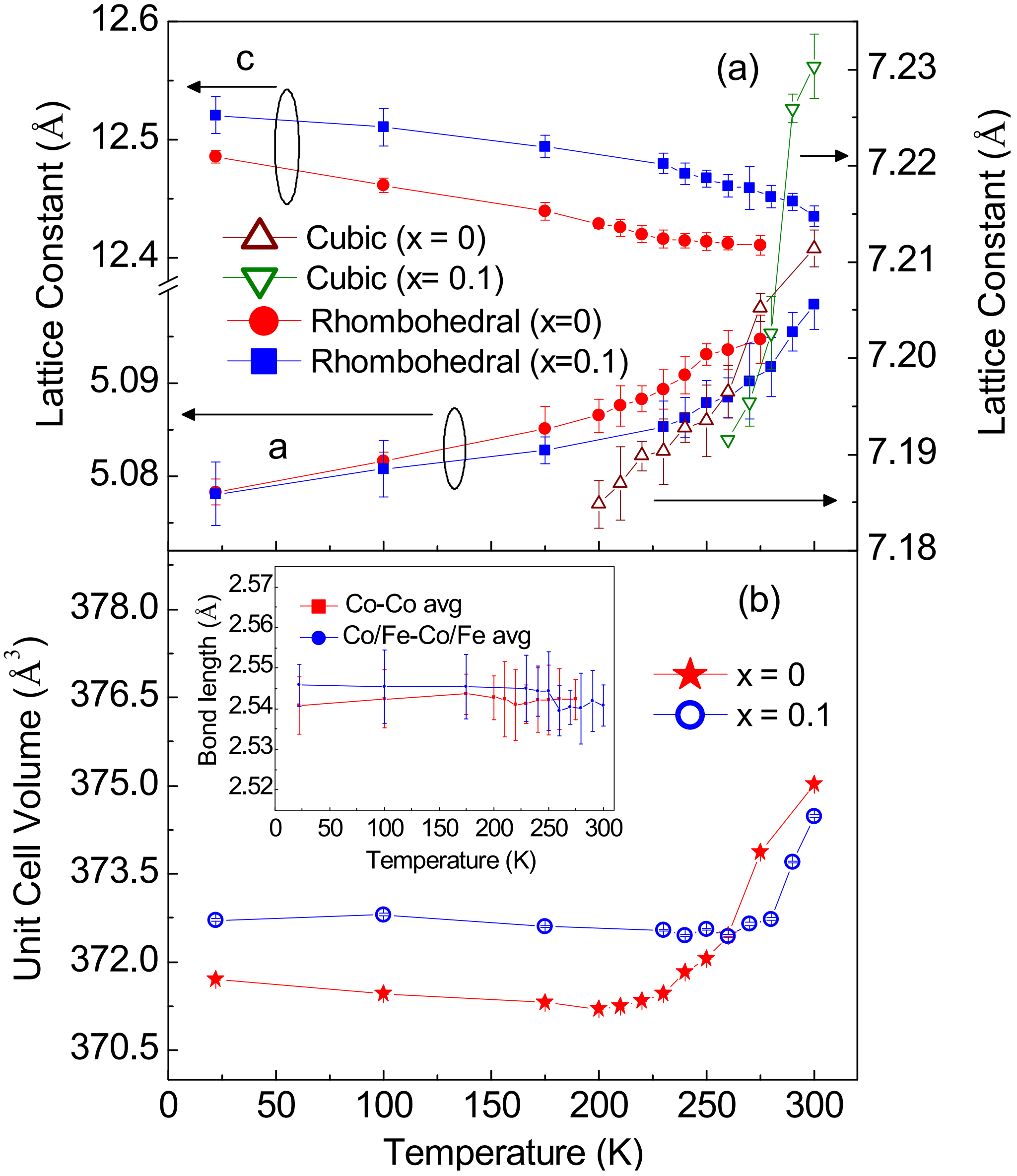

Figure 5
Halder *et al.*

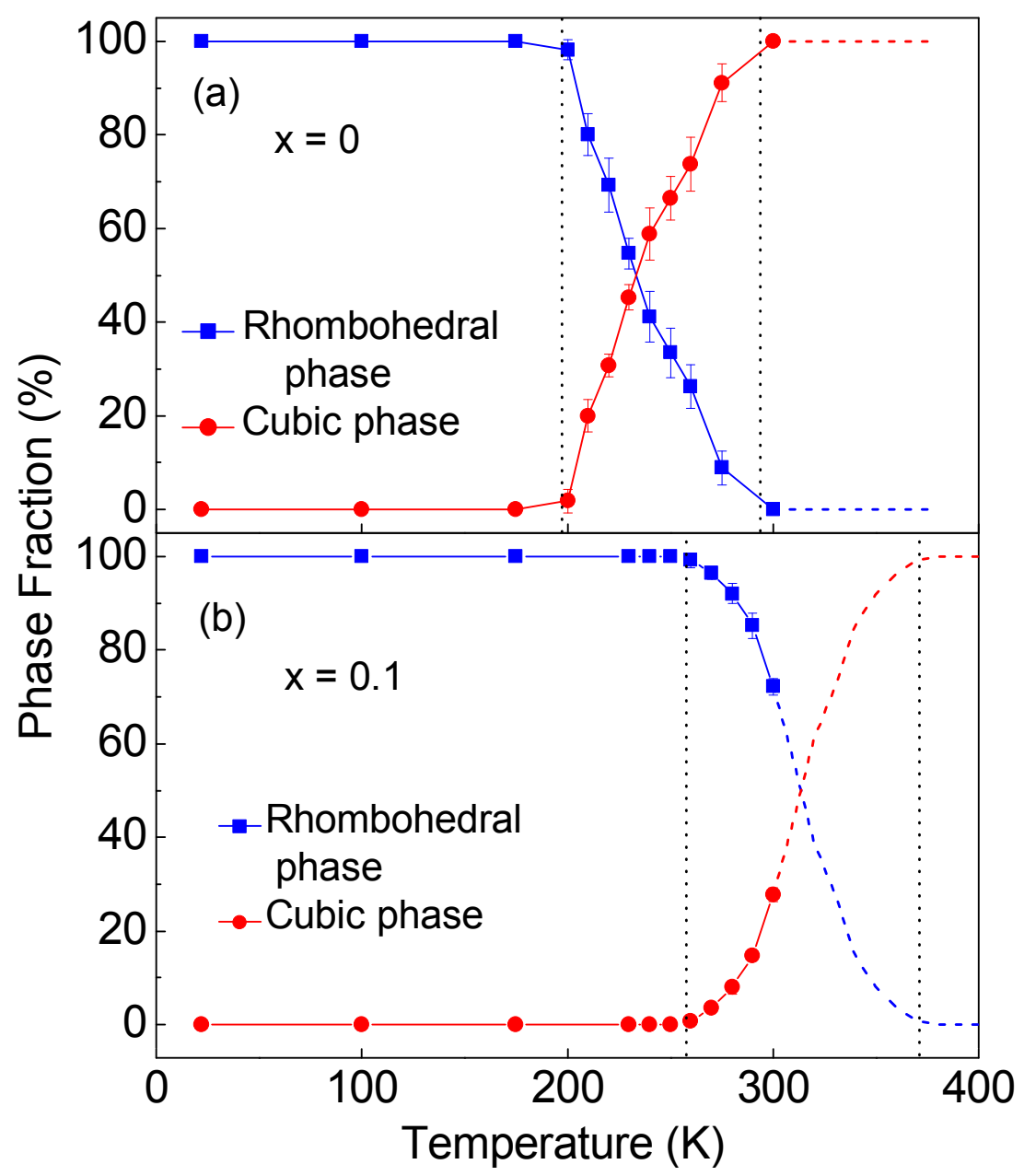

Figure 6
Halder *et al.*

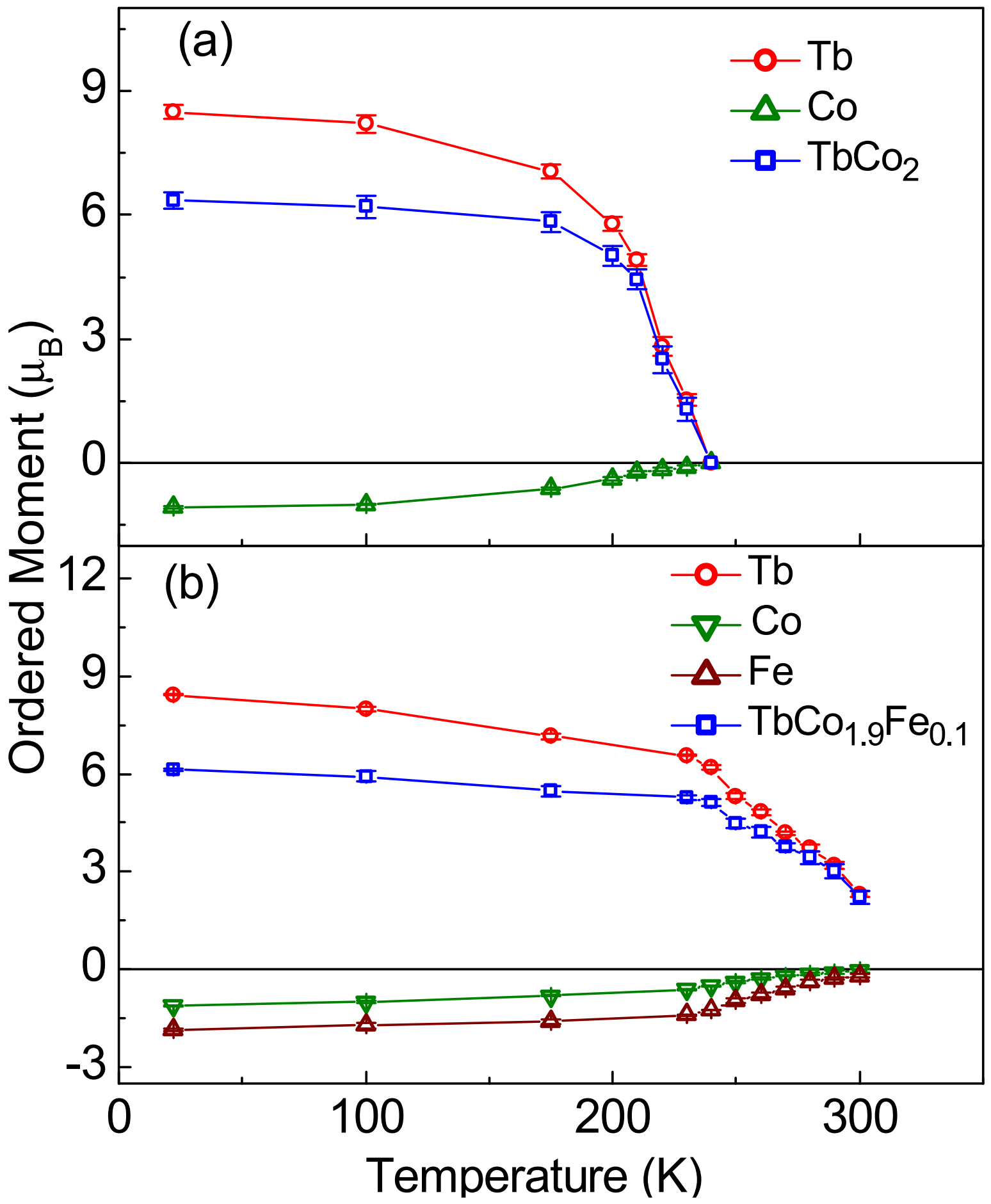

Figure 7
Halder *et al.*

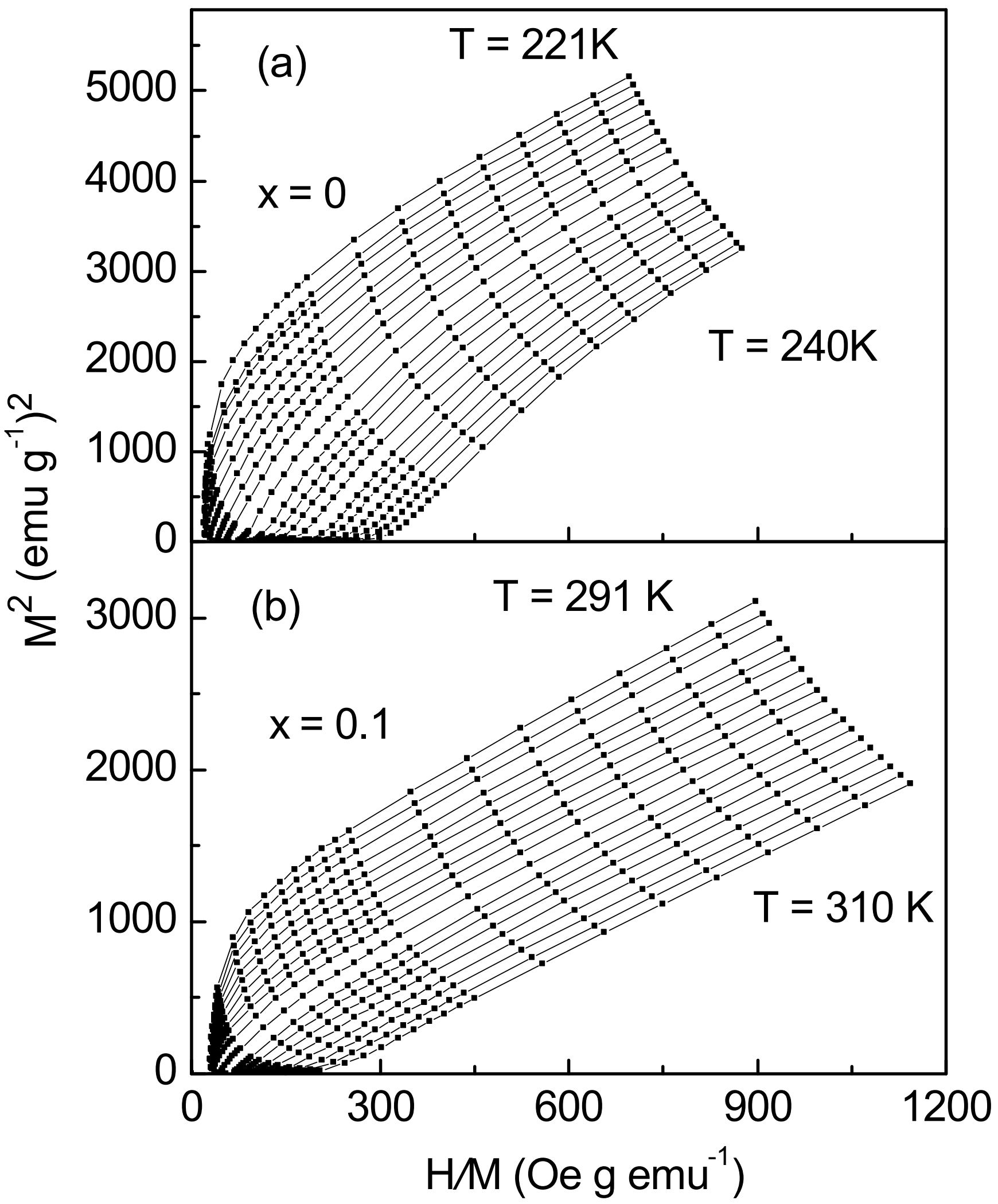

Figure 8
Halder *et al*.

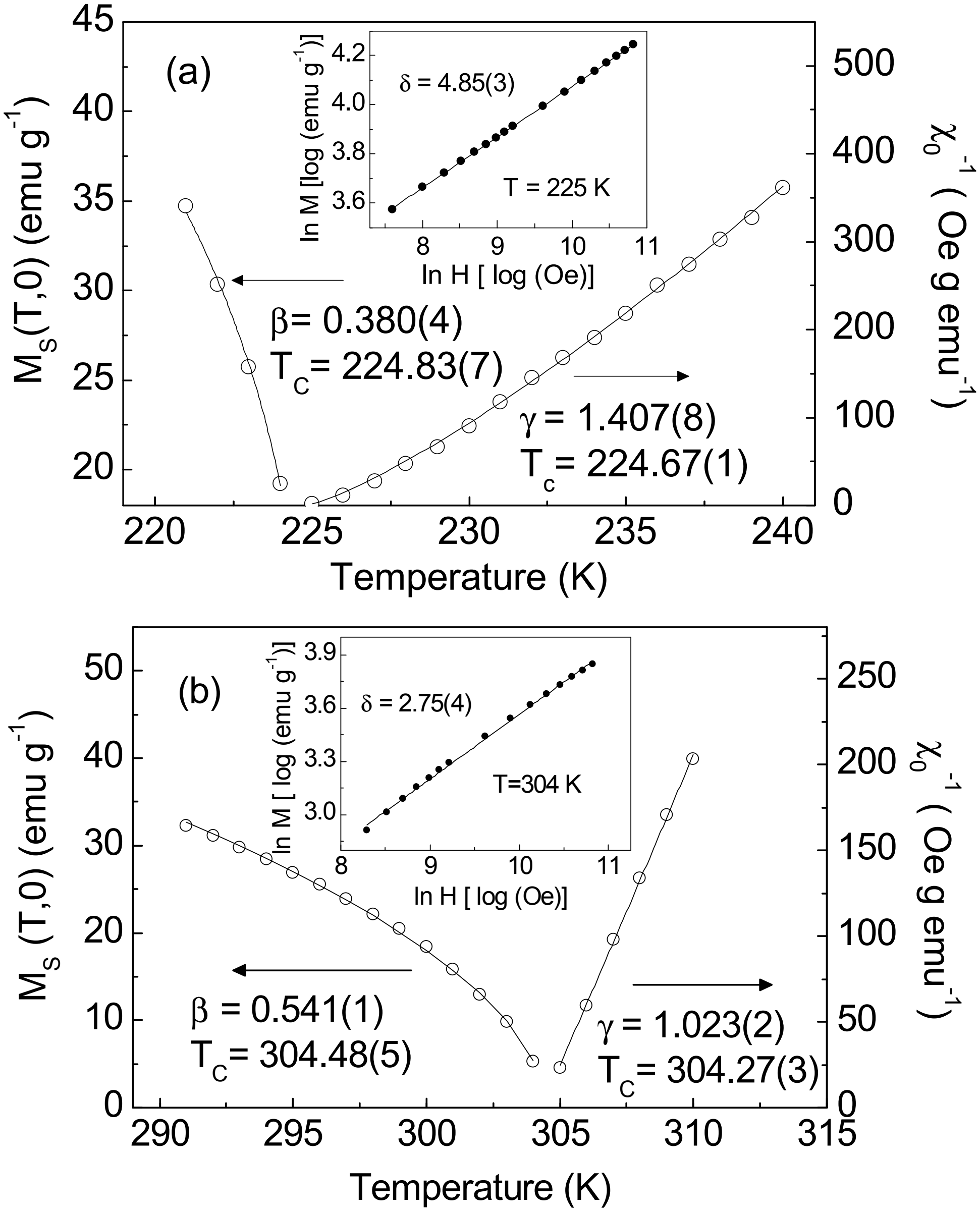

Figure 9
Halder *et al*.

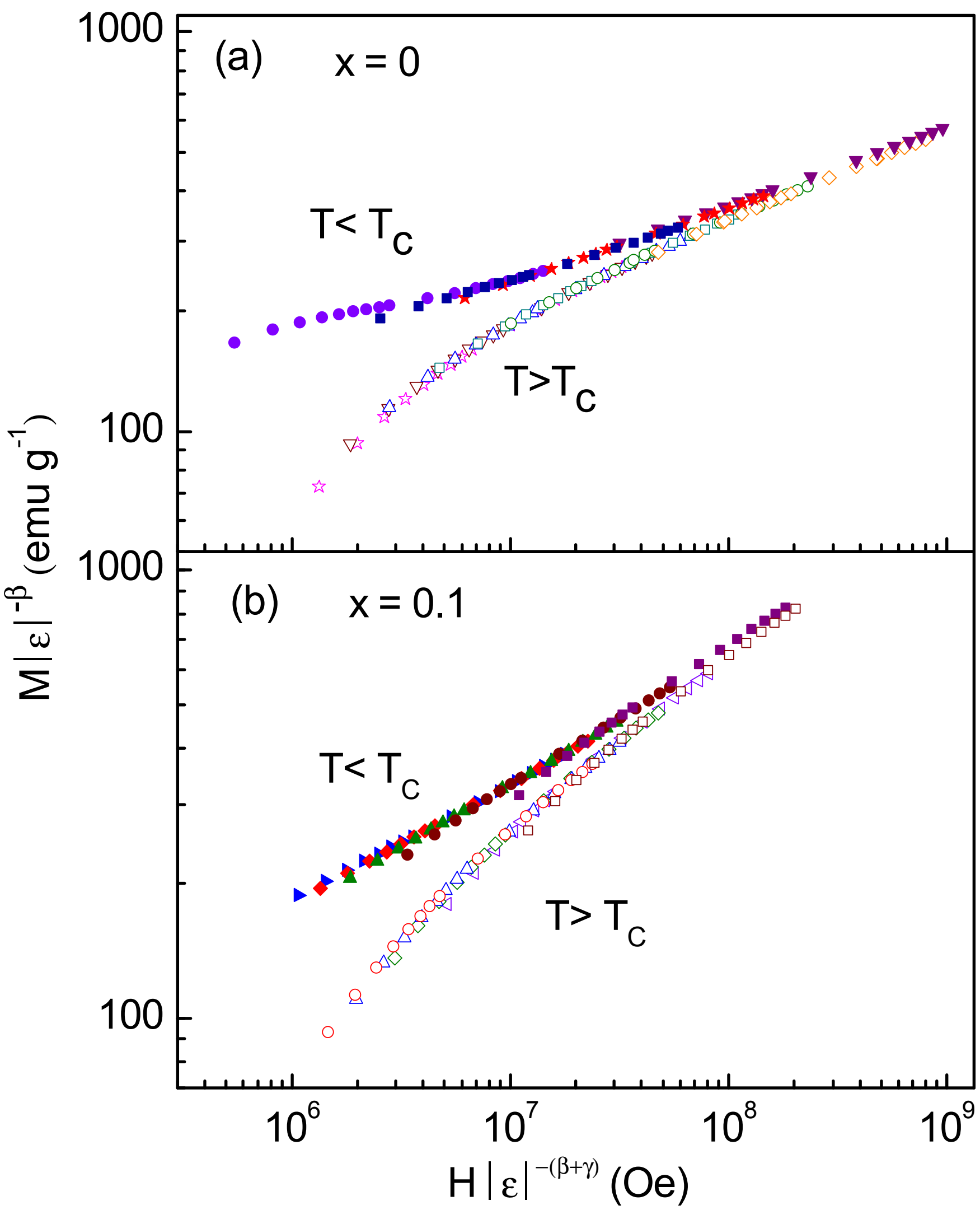

Figure 10
Halder *et al*.

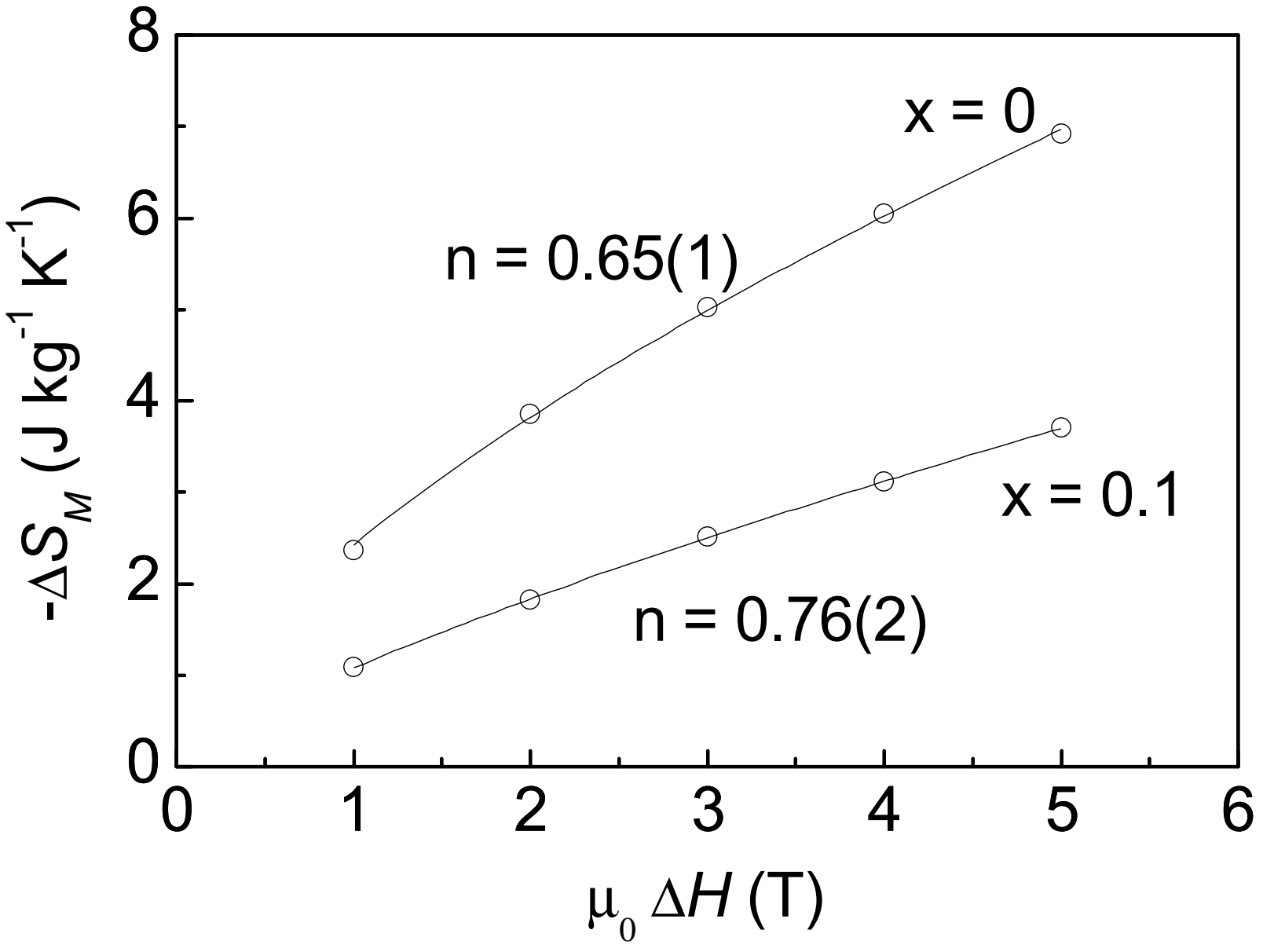

Figure 11
Halder *et al*.

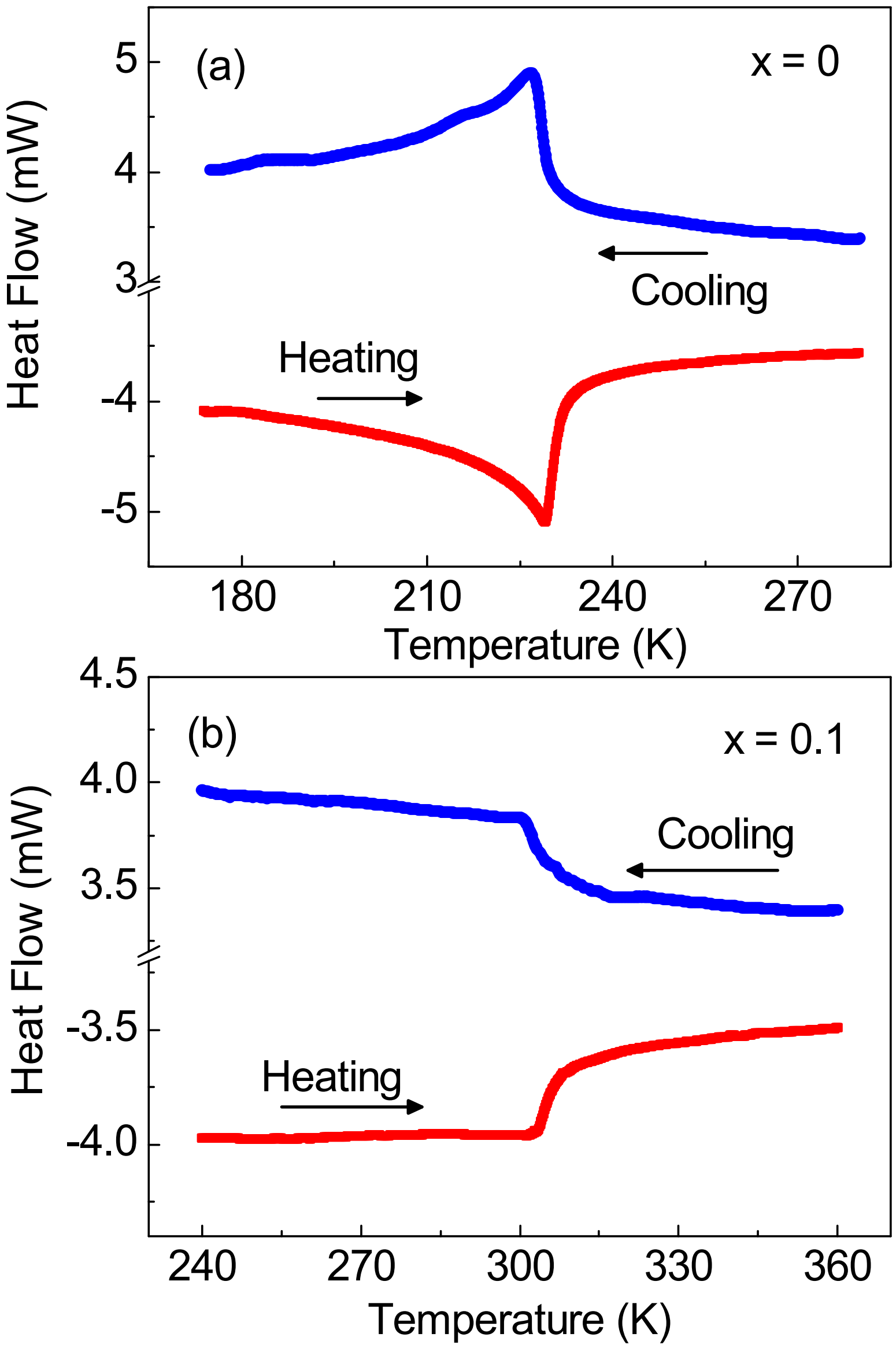

Figure 12
Halder *et al.*